# Blind speech separation based on undecimated wavelet packet-perceptual filterbanks and independent component analysis

Ibrahim Missaoui[1], Zied Lachiri[1,2]

[1] National School of Engineers of Tunis
BP. 37 Le Belvédère, 1002 Tunis, Tunisia

[2] National Institute of Applied Science and Technology
BP. 676 centre urbain cedex  Tunis, Tunisia

**Abstract**
In this paper, we address the problem of blind separation of speech mixtures. We propose a new blind speech separation system, which integrates a perceptual filterbank and independent component analysis (ICA) and using kurtosis criterion. The perceptual filterbank was designed by adjusting undecimated wavelet packet decomposition (UWPD) tree in order to accord to critical band characteristics of psycho-acoustic model. Our proposed technique consists on transforming the observations signals into an adequate representation using UWPD and Kurtosis maximization criterion in a new preprocessing step in order to increase the non-Gaussianity which is a pre-requirement for ICA.
Experiments were carried out with the instantaneous mixture of two speech sources using two sensors. The obtained results show that the proposed method gives a considerable improvement when compared with FastICA and other techniques.
*Keywords:* Perceptual Filter-Bank, Undecimated Wavelet Packet Decomposition, Independent Component Analysis, Blind speech separation.

## 1. Introduction

The blind source separation has become an interesting research topic in speech signal processing. It is a recent technique which provides one of the feasible solutions for recover the speech signals from their mixture signals without exploring any knowledge about the source signals and the mixing channel. This challenging research problem has been investigated by many researchers in the last decades, who have proposed many methods and it has been applied in various subjects including speech processing, image enhancement, and biomedical signal processing [1], [ 4].
Independent Component Analysis (ICA) is one of the popular BSS methods and often used inherently with them. It is a statistical and computational technique in which the goal is to find a linear projection of the data that the source signals or components are statistically independent or as independent as possible [1].
There are many algorithms which have been developed, using ICA method, to address the problem of instantaneous blind separation [3] such as approaches based on the mutual information minimization [9], [28], maximization of non-Gaussianity [1], [12], [10] and maximization of likelihood [9], [20]. Among these approaches, SOBI algorithm [13] is the second order blind identification which consists to diagonalize a set of covariance matrix and Jade algorithm [14] based on higher order statistics and seek to achieve the separation of the signals by using a Jacobi technique in order to performed a joint diagonalization of the cumulant matrices.
Some researchers aim to improve the performance of BSS system by combining the ICA algorithm with other techniques. For example, the approach developed in [25] combines binary time-frequency masking technique inspired from computational auditory scene analysis system [2] with ICA algorithm. Others techniques decomposes the observed signals using for example subband decomposition [26] or discrete wavelet transform [11] and then apply the separation step in each sub band. In [27], [29], a preprocessing step is employed in wavelet domain but the separation is done in time domain. The idea behind employing wavelet transformation as a preprocessing step is to improve the non-Gaussianity distribution of independent components that is a pre-requirement for ICA and to increase their independency [27], [29], [24]. Inspired from this idea, we propose a new blind separation system, in the instantaneous mixture case, to extract the speech signals of two-speakers from two speech mixtures. The proposed technique uses a perceptual filterbank which is designed by adjusting undecimated wavelet packet decomposition (UWPD) tree, according to critical band characteristics of psycho-acoustic model [15], for the transformation of the two mixtures signals into adequate representation to emphasize the non-Gaussian nature of mixture signals.

This paper is organized as follows. Section 2 introduces the blind speech separation problem and describes the FastICA algorithm. Section 3 presents the principle of the undecimated wavelet packet decomposition and perceptual filterbank. Then in section 4, the proposed method is described. Section 5 exposes the experimental results. Finally, Section 6 concludes and gives a perspective of our work.

## 2. Blind Speech Separation

2.1 Problem Statement

The objective of Blind Speech Separation is to extract the original speech signals from their observed mixtures without reference to any prior information on the sources signals or the observed mixtures. The latter contain a different combination of the source signals and can be mathematically described by:

$$X(t) = AS(t) \qquad (1)$$

Where $X(t)=[x_1(t)…x_n(t)]^T$ is a vector of mixture signals, $S(t)=[s_1(t)…s_m(t)]^T$ is the unknown vector of sources signals and A is the unknown mixing matrix having dimension (m*n).
Independent Component Analysis is a typical BSS method which tends to solve this problem. The purpose of the ICA is to find a separating matrix or an unmixing matrix $W=A^{-1}$, which is used to calculate the estimated signal $S(t)$ of source signals as $S(t)=WX(t)$. To estimate W, we have to make some fundamental assumptions and impose certain restrictions [1]: The components $s_i(t)$ of S(t) (i.e. the sources) are assumed to be statistically independent with non-gaussian distribution.

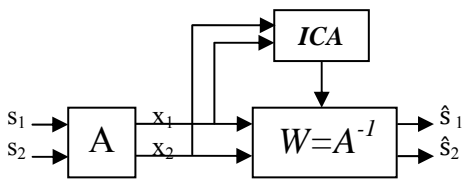

Fig. 1 Principle of ICA.

In other words, ICA can be defined as a method that researches a linear transformation, which maximizes the non-Gaussianity of the components of S(t). To measure the non gaussianity, kurtosis or differential entropy called negentropy can be employed. FastICA algorithm [12], [1], [8] is one of the most popular algorithms performing independent component analysis. The Principle of ICA can be depicted as in Figure 1.

2.2 FastICA Method

The FastICA algorithm (A Fast Fixed-Point algorithm of independent component analysis) is a technique proposed and developed by Aapo Hyvarinen and al [1], which is characterized by a high order convergence rate. In this approach, the separation task is based on a point iteration scheme in order to find the maximum of the non-Gaussianity of a projected component. The non-Gaussianity, which is the function of contrast of FastICA algorithm, can be measured with the differential entropy, as known as negentropy [12]. The latter is defined as the difference between the entropy of a Gaussian random vector $y_{gauss}$ of same covariance matrix as y and the random vector y:

$$J(y) = H(y_{gauss}) - H(y) \qquad (2)$$

Where H(y) is the differential entropy of y and it is computed as follows:

$$H(y) = -\int f(y)\log(f(y))dy \qquad (3)$$

The negentropy can be considered as the optimal measure of the non gaussianity. However, it is difficult to estimate the true negentropy. Thus, several approximations are used and developed such the one developed by Aapo Hyvarinen et al [1], [12]:

$$J(y) = \sum_{i=1}^{p} k_i \left( E[g_i(y)] - E[g_i(\upsilon)] \right)^2 \qquad (4)$$

Where $k_i$, $g_i$ and $\nu$ are respectively positive constants, the non quadratic functions and Gaussian random variable.
The separating matrix W is calculated using a fundamental fixed-point iteration which performed by using the following expression:

$$W_i(k) = E\left\{\hat{X}_i g(W_i^T \hat{X}_i)\right\} - E\left\{g'(W_i^T \hat{X}_i)\right\} W_i \qquad (5)$$

## 3. Undecimated Wavelet Packet-Perceptual Filterbank

3.1 Wavelet Transform

Wavelet Transform [5], [18], represents an alternative technique for the processing of non-stationary signals which provides a linear powerful representation of signals. The discrete wavelet transforms (DWT) is a multi-resolution representation of a signal which decomposes signals into basis functions. It is characterized by a higher time resolution for high frequency components and a higher frequency resolution for low frequency components. The DWT consists on filtering the input signal by two filters H (a low-pass filter) and G (a high-pass filter), leading two sub-bands called respectively approximations and details, followed by a decimation factor of two. This filtering process is then iterated only for the approximation sub-band at each level of decomposition [6].

The wavelet packet decomposition (WPD), viewed as a generalization of the discrete wavelet transform (DWT), aims to have a more complete interpretation of the signal, in which the filtering process is applied to decompose on both approximations and details sub-bands and still decimates the filters outputs [7].

To provide a denser approximation and to preserve the translation invariance, the undecimated wavelet packet transform (UWPT) has been introduced and was invented several times with different names as algorithm à trous (algorithm with holes) [17], shift invariant DWT [22] and redundant wavelet transform [16]. The UWPT is computed in a similar manner as the wavelet packet transform except that the downsampling operation after each filtering step is suppressed.

### 3.2 Perceptual filterbank

In the proposed blind speech separation system, we use a perceptual filterbank which is designed using undecimated wavelet packet decomposition [15]. The decomposition tree consists on five levels full UWPD tree using Daubechies 4 (db4) of an 8 kHz speech signal. This decomposition tree structure is adjusted in order to accord to critical band characteristics. The result tree was called critical bands-undecimated wavelet package decomposition (CB-UWPD) tree. Indeed, the audible frequency range of human auditory is 20 to 20000 Hz which can be approximated with 25 barks. However, the sampling frequency chosen is 8 kHz leading to a bandwidth of 4 kHz. As shown in table 1, this bandwidth containes approximately 17 critical bands (barks). The tree structure of CB-UWPD obtained according to the results critical bandwidths (CBW) is depicted in fig 1.

The following equation for each node of the tree is given the corresponding to the critical bandwidths (CBW):

$$cbw(i,j) = 2^{-j}(F_s - 1) \qquad (6)$$

Where i=(0,1,..,5) and j=(0.., $2^{-j}$-1) are respectively the number of levels and the position of the node and $F_s$ is the sampling frequency.

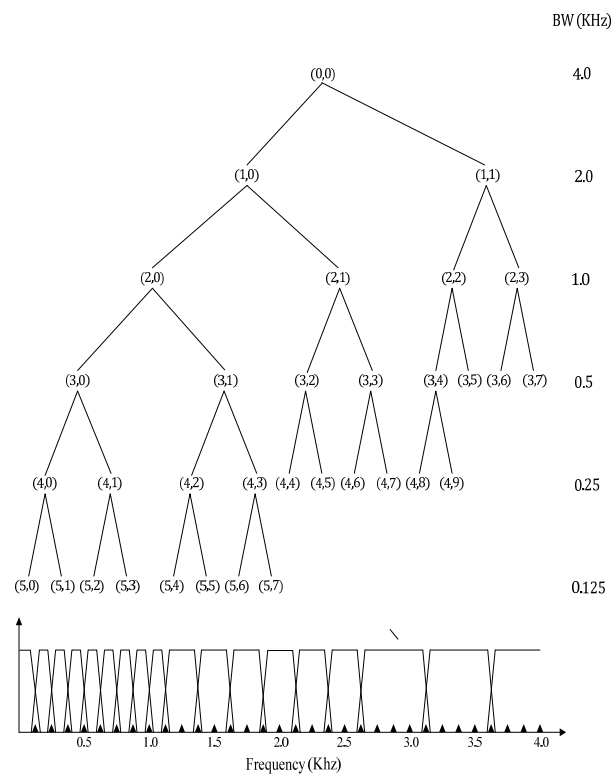

Fig. 1 The CB-UWPD tree and its corresponding frequency bandwidths(perceptual filterbank).

Table 1: Critical Band Characteristics

| Critical bands (barks) | Center frequency (Hz) | Critical bandwidth (CBW) (Hz) |
|---|---|---|
| 1 | 50 | 100 |
| 2 | 150 | 100 |
| 3 | 250 | 100 |
| 4 | 350 | 100 |
| 5 | 450 | 110 |
| 6 | 570 | 120 |
| 7 | 700 | 140 |
| 8 | 840 | 150 |
| 9 | 1000 | 160 |
| 10 | 1170 | 190 |
| 11 | 1370 | 210 |
| 12 | 1600 | 240 |
| 13 | 1850 | 280 |
| 14 | 2150 | 320 |
| 15 | 2500 | 380 |
| 16 | 2900 | 450 |
| 17 | 3400 | 550 |

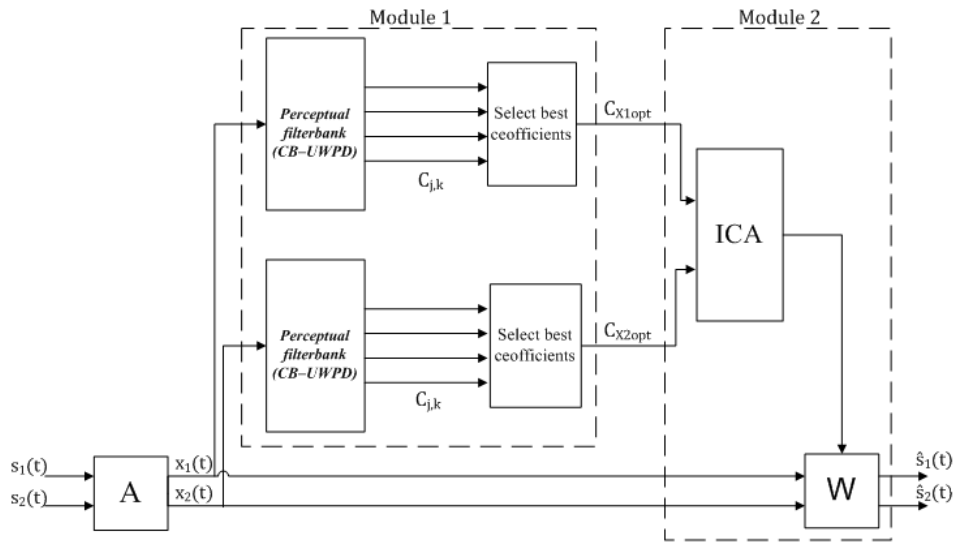

Fig. 2 The framework of proposed speech separation system

## 4. The proposed Method

We suggest extracting the speech signals of two speakers from two speech mixtures. The proposed speech separation system, as depicted in Figure 1, contains two modules shown in dotted boxes. In the first module, the speech mixtures $x_1(n)$ and $x_2(n)$ are processed by a perceptual filterbank which designed by adjusting undecimated wavelet packet decomposition (UWPD) tree, according to critical bands of psycho-acoustic model of human auditory system. In order to increase the non Gaussianity that is a pre-requirement for ICA, we select the appropriate coefficients of the two mixtures which having the high non-Gaussian nature of distribution. The two result signals are then used as two new inputs of the second module. The latter performs the source separation using FastICA algorithm. The description of each module is given in the following sub-sections.

### 4.1 Preprocessing Module

In this section, we explain the preprocessing module that decomposes the observed signals by perceptual filterbank. This filterbank is designed by adjusting undecimated wavelet packet decomposition tree to accord critical band characteristics of psycho-acoustic model [15]. Each result coefficients of the two mixtures $x_1(n)$ and $x_2(n)$ can be viewed as an appropriate signal. In order to increase the non Gaussianity of the signals that is a pre-requirement for ICA, we need to find the best coefficients of each mixture which have the highest non-Gaussian nature of distribution. Thus, the performance of source separation task will be improved. The selection of the best coefficients can be performed using Shannon entropy criterion [27], [29]. In our case, we chose to use the kurtosis (forth order cumulant) as a criterion to select the best coefficients instead of Shannon entropy criterion. The procedure of the selection algorithm is give as follows:

Step 1: Decompose the mixture signals into undecimated wavelet packet.
Step 2: Calculate the kurtosis of each node $C_{j,k}$ of UWPD tree.
Step 3: Select the node which has the highest kurtosis.

The Kurtosis (forth order cumulant) for each node can be estimated by using the fourth moment. It is defined as follows:

$$kurt(y(i)) = E\left[y(i)^4\right] - 3\left(E\left[y(i)^2\right]\right)^2 \qquad (7)$$

Where $y(i)$ is a vector of UWPD coefficients at each node. We assume that $y(i)$ is zero-mean and have unit energy.

The forth order cumulant (Kurtosis) represents the classical measure of non gaussianity of signal [1]. Therefore seek to maximize the kurtosis correspond to find the representation of signal which own high non-Gaussian nature of distribution. Consequently, during the application of ICA that exploits the non-Gaussianity in separation task, we will have a significant gain.

### 4.2 Separation Module

This module is the separation module. It can be devised into two steps. The first step consists of generating the unmixing matrix W using the FastICA algorithm. The select UWPD coefficients of two mixtures signals $x_1(n)$ and $x_2(n)$ which obtained in preprocessing module are used as two inputs signals of FastICA algorithm. In the second step, the separated signals are obtained by taking into account the original mixtures signals.

## 5. Results and Evaluation

To evaluate the performance of the proposed blind speech separation Method, described in section 4. We use some sentences taken from TIMIT database, this database consists of speech signals of a total of 6300 sentences formed by 10 sentences spoken by each of 630 speakers from 8 major dialect regions of the United States [23]. We consider two speech mixtures composed of two speakers, so we mixes in instantaneous two speech signals, which are respectively pronounced by male and female speaker, two female speakers and two male speakers. The two speech mixtures are generating, artificially, using mixing matrix as:

$$A = \begin{pmatrix} 2 & 1 \\ 1 & 1 \end{pmatrix} \quad (8)$$

The performance evaluation of our work includes different performance metrics such as the blind separation performance measures used in BSS EVAL [19], [30], including the signal to interference ratio SIR and the signal to distortion ratio SDR measures. The principle of these measures consists on decomposing the estimated signal $s_i(n)$ into the following component sum:

$$s_i(n) = s_{target}(n) + s_{interf}(n) + s_{artefact}(n) \quad (9)$$

where $s_{target}(n)$, $e_{interf}(n)$ and $e_{artefact}(n)$ are, respectively, an allowed deformation of the target source $s_i(n)$ an allowed deformation of the sources which takes account of the interference of the unwanted sources and an artifact term which represents the artifacts produced by the separation algorithm. The two performance criteria SIR and SDR are computed using the last decomposition as following:

$$SIR = 20 \log \frac{\|s_{target}(n)\|^2}{\|s_{interf}(n)\|^2} \quad (10)$$

$$SDR = 20 \log \frac{\|s_{target}(n)\|^2}{\|s_{interf}(n)\|^2 + \|s_{artefact}(n)\|^2} \quad (11)$$

In addition, the recovered speech signals are evaluated with the segmental, overall signal to noise ratio (SNR) and the Perceptual Evaluation of Speech Quality(PESQ). The PESQ is defined in the ITU-T P.862 standard [21] and represents an objective method for evaluating the speech quality. The resulting of PESQ measurement is equivalent to the subjective "Mean Opinion Score" (MOS) measured score.

In the previous experiments, we compare our system with FastICA algorithm [12] and two well-known algorithms Jade [14] and SOBI [13].

The experimental results are shown in three tables which reports the evaluation measures obtained for three example cases of mixture signal. Table 2 lists the separate performance measures including ratio SIR and SDR obtained after separation by Sobi, Jade, FastICA and the proposed method. We observed that the SIR≈SDR and their values is better for the proposed method than that of FastICA, jade and SOBI in the majority of cases for the two signals. The SIR average where we have a mixture composed with two female speakers (or experiment 2) for exemple, is 14.06 for SOBI, 43.12 db for Jade, 39.80 for FastICA and 45.10 db for proposed method. The improvement in the SIR and SDR ratio average is particularly significant in the case of mixture observed formed by two male speaker signals. The improvement average in this case between the proposed method and FastICA is 15.45 db.

Table 3 and table 4 shows that the estimated signals obtained by using the proposed method is better than those obtained by FastICA and the two algorithms Jade and SOBI for the three experiments. We have obtained, for exemple, seg SNR egale to 33.90 db using proposed method and 29.14 db using FastICA.

In order to have a better idea about the quality of estimated signal obtained, PESQ has been used. It is regarded as one of the reliable methods of subjective test. It returns a score from 0.5 to 4.5. Table 5 illustrates the PESQ score obtained. We see that the proposed method is still more effective in terms of perceptual quality than FastICA, jade and SOBI.

Table 2: Comparison of SIR and SDR using SOBI, Jade, Fast-ICA and proposed Method (PM)

| | | SOBI | Jade | FastICA | PM |
|---|---|---|---|---|---|
| Experiment 1 Female (F)+Male (M) | SIR (F.speaker) | 26.92 | 54.72 | 44.39 | 51.11 |
| | SIR (M.speaker) | 26.29 | 45.63 | 51.68 | 60.75 |
| | SDR (F.speaker) | 26.92 | 54.72 | 44.39 | 51.11 |
| | SDR (M.speaker) | 26.29 | 45.63 | 51.68 | 60.75 |
| | Average | 26.60 | 50.17 | 48.03 | 55.93 |
| Experiment 2 Female (F)+ Female (F) | SIR (F.speaker 1) | 14.39 | 41.37 | 44.57 | 51.62 |
| | SIR (F.speaker 2) | 13.74 | 44.87 | 35.04 | 38.59 |
| | SDR (F.speaker 1) | 14.39 | 41.37 | 44.57 | 51.62 |
| | SDR (F.speaker 2) | 13.74 | 44.87 | 35.04 | 38.59 |
| | Average | 14.06 | 43.12 | 39.80 | 45.10 |
| Experiment 3 Male (M)+Male (M) | SIR (M.speaker 1) | 18.46 | 65.02 | 46.20 | 72.22 |
| | SIR (M.speaker 2) | 19.57 | 37.32 | 48.37 | 53.25 |
| | SDR (M.speaker 1) | 18.46 | 65.02 | 46.20 | 72.22 |
| | SDR (M.speaker 2) | 19.57 | 37.32 | 48.37 | 53.25 |
| | Average | 19.01 | 51.17 | 47.28 | 62.73 |

Table 3: Comparison of segmental SNR using SOBI, Jade, FastICA and proposed Method (PM)

| | | SOBI | Jade | FastICA | PM |
|---|---|---|---|---|---|
| Experiment 1 Female (F)+Male (M) | Seg SNR (F.speaker) | 22.58 | 33.56 | 30.79 | 32.79 |
| | Seg SNR (M.speaker) | 20.47 | 29.40 | 31.15 | 33.03 |
| Experiment 2 Female (F)+ Female (F) | Seg SNR (F.speaker 1) | 15.19 | 32.01 | 32.73 | 33.76 |
| | Seg SNR (F.speaker 2) | 12.27 | 32.12 | 28.37 | 30.07 |
| Experiment 3 Male (M)+Male (M) | Seg SNR (F.speaker 1) | 13.47 | 33.20 | 29.14 | 33.90 |
| | Seg SNR (F.speaker 2) | 20.89 | 30.88 | 33.56 | 34.10 |

Table 4: Comparison of overall SNR using SOBI, Jade, FastICA and proposed Method (PM)

| | | SOBI | Jade | FastICA | PM |
|---|---|---|---|---|---|
| Experiment 1 Female (F)+Male (M) | Overall SNR (F.speaker) | 26.92 | 54.72 | 44.39 | 51.11 |
| | Overall SNR (M.speaker) | 26.29 | 45.63 | 51.68 | 60.75 |
| Experiment 2 Female (F)+ Female (F) | Overall SNR (F.speaker 1) | 14.37 | 41.37 | 44.57 | 51.62 |
| | Overall SNR (F.speaker 2) | 13.82 | 44.87 | 35.04 | 38.59 |
| Experiment 3 Male (M)+Male (M) | Overall SNR (F.speaker 1) | 18.47 | 37.32 | 46.20 | 72.22 |
| | Overall SNR (F.speaker 2) | 19.55 | 30.88 | 48.37 | 53.25 |

Table 5: Comparison of PESQ using SOBI, Jade, FastICA and proposed Method (PM)

| | | SOBI | Jade | FastICA | PM |
|---|---|---|---|---|---|
| Experiment 1 Female (F)+Male (M) | PESQ (F.speaker) | 2.58 | 3.29 | 3.25 | 3.29 |
| | PESQ (M.speaker) | 3.45 | 4.14 | 4.27 | 4.38 |
| Experiment 2 Female (F)+ Female (F) | PESQ (F.speaker 1) | 1.53 | 4.20 | 4.27 | 4.42 |
| | PESQ (F.speaker 2) | 0.88 | 3.65 | 3.40 | 3.52 |
| Experiment 3 Male (M)+Male (M) | PESQ (F.speaker 1) | 1.53 | 2.24 | 2.06 | 2.24 |
| | PESQ (F.speaker 2) | 1.20 | 4.23 | 4.42 | 4.47 |

## 6. Conclusions

In this paper, we proposed a new blind speech separation system in the instantaneous case. This system consists on a combination of ICA algorithm with undecimated wavelet packet transform. The latter is used as a preprocessing module using a Kurtosis maximization criterion in order to increase the non-Gaussian nature of the signals. The results signals are then employed to perform a preliminary separation leading to the inverse matrix W used to separate the signals in the time domain. The experimental results show that the proposed approach yield to a better separation performance compared to FastICA and two well-known algorithms.

For future work, we aim to separate the convolutive mixtures with the proposed system.

## References

[1] A. Hyvärinen, J. Karhunen, and E. Oja, Independent Component Analysis, New York: Wiley-Interscience, 2001.
[2] L. Wang, and G. J. Brown, Computational Auditory Scene Analysis: Principles, Algorithms, and Applications, Hoboken NJ :Wiley/IEEE Press, 2006.
[3] S. Haykin, Neural Networks and Learning Machines (third ed.), Prentice-Hall, 2008.
[4] A. Cichocki, and S. Amari, Adaptive Blind Signal and Adaptive Blind Signal and Image Processing. New York: John Wiley and Sons, 2002.
[5] C.S. Burrus, R.A. Gopinath, and H. Guo, Introduction to Wavelets and Wavelet Transformr: A Primer, Prentice Hall, 1998.
[6] S. Mallat, A Wavelet Tour of Signal Processing: The Sparse Way, 3rd ed, London: Academic Press, 2008.
[7] R. R. Coifman, Y. Meyer, and M. V. Wickerhauser, "Wavelet analysis and signal processing, in Wavelets and their applications. Boston, MA: Jones and Bartlett, pp.153-178, 1992.
[8] P. Comon, "Independent components analysis: A new concept?", Signal Processing, Vol. 36, No. 3, 1994, pp. 287-314.
[9] A.J. Bell and T.J. Sejnowski, "An information maximization approach to blind separation and blind deconvolution", Neural Computation, Vol. 7, 1995, pp. 1004-1034.
[10] F. S. Wang, H. W. Li, R. Li, "Novel Non Gaussianity Measure Based BSS Algorithm for Dependent Signals", Lecture Notes in Computer Science, Vol. 4505, 2007, pp. 837-844.
[11] W. Xiao, H. Jingjing, J. Shijiu, X. Antao, and W. Weikui, "Blind separation of speech signals based on wavelet transform and independent component analysis", Transactions of Tianjin University, Vol. 16, No. 2, 2010, pp 123-128.
[12] A. Hyvrinen, "Fast and robust fixed-point algorithms for independent component analysis". IEEE Transactions on Neural Networks, Vol. 10, No.3, 1999, pp. 626-634.
[13] A. Belouchrani, K. Abed-Meraim, J.-F. Cardoso, and E. Moulines, A blind source separation technique using second order statistics, IEEE Transactions on Signal Processing, Vol. 45, 1997, pp. 434-444.
[14] J.F. Cardoso, "Higher-order contrasts for independent component analysis", Neural Computation, Vol. 11, 1999, pp.157-192.
[15] H. Tasmaz, and E. Ercelebi, "Speech enhancement based on undecimated wavelet packet-perceptual filterbanks and MMSE-STSA estimation in various noise environments", Digital Signal process, Vol. 18, No. 5, 2008, pp. 797-812.
[16] J. Fowler, "The redundant discrete wavelet transform and additive noise", IEEE Signal Processing Letters, Vol. 12, No. 9, 2005, pp. 629-632.
[17] M. Shensa, "The discrete wavelet transform: Wedding the à trous and Mallat algorithms", IEEE Transactions on Signal Processing, Vol. 40, No. 10, 1992, pp. 2464-2482.
[18] C. Gargour, M. abrea, , V. Ramachandran, and J.M. Lina, "A short introduction to wavelets and their applications", IEEE Circuits and Systems Magazine, Vol. 9, No. 2, 2009, pp. 57-58.
[19] E. Vincent, R. Gribonval, and C. Fevotte, "Performance Measurement in Blind Audio Source Separation", IEEE Transactions on Audio, Speech, and Language Processing, Vol. 14, No. 4, 2006, pp 1462-1469.
[20] J. T. Chien, B. C. Chen, "A New Independent Component Analysis for Speech Recognition and Separation", IEEE transactions on audio, speech and language processing, Vol. 14, No. 4,2006, pp. 1245 - 1254.
[21] ITU-T P.862, "Perceptual evaluation of speech quality (PESQ), an objective method for end-to-end speech quality assessment of narrow-band telephone networks and speech codecs", International Telecommunication Union, Geneva, 2001.
[22] A. T. Walden, and C. Contreras, "The phase-corrected undecimated discrete wavelet packet transform and its application to interpreting the timing of events", in Proceedings of the Royal Society of London, 1998, Vol. 454, pp. 2243-2266.
[23] W. Fisher, G. Dodington, and K. Goudie-Marshall. "The TIMIT-DARPA speech recognition research database: Specification and status", In. DARPA Workshop on Speech Recognition, 1986.
[24] K. Usman, H. Juzoji, I. Nakajima, and M.A. Sadiq, "A study of increasing the speed of the independent component analysis (lCA) using wavelet technique", in Proceedings of International Workshop on Enterprise Networking and Computing in Healthcare Industry (HEAL THCOM 2004), 2004, pp. 73-75.
[25] M.S. Pedersen, D.L. Wang, J. Larsen, and U. Kjems "Overcomplete blind source separation by combining ICA and binary time-frequency masking", in Proceedings of IEEE Workshop on Machine Learning for Signal Processing, 2005, pp. 15-20.
[26] T. Tanaka and A. Cichocki, "Subband decomposition independent component analysis and new performance criteria", in Proceedings of the IEEE Conference on Acoustics, Speech and Signal Processing, 2004, pp. 541-544.
[27] R. Moussaoui, J. Rouat and R. Lefebvre, "Wavelet Based Independent Component Analysis for Multi-Channel Source Separation", in Proceedings of IEEE International Conference on Acoustics, Speech and Signal Processing, 2006, Vol. 5, pp. 645-648.

[28] J. T. Chien, H. L. Hsieh, and S. Furui, "A new mutual information measure for independent component alaysis", in Proceedings of IEEE International Conference on Acoustics, Speech and Signal Processing, 2008, 1817-1820,.

[29] M.R. Mirarab, M.A Sobhani, and AA Nasiri, "A New Wavelet Based Blind Audio Source Separation Using Kurtosis", in International Conference on Advanced Computer Theory and Engineering, 2010, Vol. 4, pp. 36-39.

[30] C. Fevotte, R. Gribonval, and E. Vincent, BSS EVAL toolbox user guide, IRISA, Rennes, France, Technical Report 1706, 2005.

**Ibrahim Missaoui** was born in Tunisia. He received his M.S degree in automatic and Signal processing from the National School of engineering of Tunis (ENIT) in 2007. He started preparing his Ph.D degree in Electrical Engineering in 2008. Her current research area is the blind speech separation.

**Zied Lachiri** was born in Tunis, Tunisia. He received the M.S. degree in automatic and signal processing and the Phd. degree in electrical engineering from the National School of Engineer of Tunis (ENIT- Tunisia), in 1998 and 2002, respectively.
In 2000, he joined the Applied Sciences and Technology National Institute (INSAT), as research Assistant and became Assistant Professor in 2002. He is currently an Associate Professor at the Department of Physic and instrumentation (INSAT) and member of Systems and Signal Processing Laboratory (LSTS-ENIT)). His research interests include signal processing, image processing and pattern recognition, applied in biomedical, multimedia, and man machine communication. He is Member of the EURASIP, European Association for Signal, Speech and Image Processing.